# Dynamic Cell Modeling of Li-Ion Polymer Batteries for Precise SOC Estimation in Power-Needy Autonomous Electric Vehicles

Qasim Ajao, and Lanre Sadeeq, *(Members, IEEE)*

**Abstract**

This paper presents findings on dynamic cell modeling for state-of-charge (SOC) estimation in an autonomous electric vehicle (AEV). The studied cells are Lithium-Ion Polymer-based with a nominal capacity of around 8Ah, optimized for power-needy applications. The AEV operates in a harsh environment with rate requirements up to ±25C and highly dynamic rate profiles, unlike portable-electronic applications with constant power output and fractional C rates. SOC estimation methods effective in portable-electronics may not suffice for the AEV. Accurate SOC estimation necessitates a precise cell model. The proposed SOC estimation method utilizes a detailed Kalman-filtering approach. The cell model must include SOC as a state in the model state vector. Multiple cell models are presented, starting with a simple one employing "Coulomb counting" as the state equation and Shepherd's rule as the output equation, lacking prediction of cell relaxation dynamics. An improved model incorporates filter states to account for relaxation and other dynamics in closed-circuit cell voltage, yielding better performance. The best overall results are achieved with a method combining nonlinear autoregressive filtering and dynamic radial basis function networks. The paper includes lab test results comparing physical cells with model predictions. The most accurate models obtained have an RMS estimation error lower than the quantization noise floor expected in the battery-management-system design. Importantly, these models enable precise SOC estimation, allowing the vehicle controller to utilize the battery pack's full operating range without overcharging or undercharging concerns.

**Keywords:** Autonomous electric vehicle, Modeling, battery model, battery management systems (BMS), lithium polymer, state of charge, Kalman-Filter

## I.  INTRODUCTION

The focus of this paper is to outline several techniques for capturing the electrical input-output characteristics of Lithium-Ion Polymer Battery (LiPB) cells. We approach the modeling of these cells as nonlinear dynamic systems and represent them in a discrete-time state-space framework. More specifically, we adopt the following structure:

$$\begin{aligned} x_{k+1} &= f(x_k, u_k) + w_k \\ y_k &= g(x_k, u_k) + v_k, \end{aligned} \quad (1)$$

where $x_k$ represent the system state vector at discrete-time index $k$. The system input $u_k$ refers to the measured exogenous system input at time $k$ (which may include measurements of battery-pack current, temperature and so forth) and $w_k$ is unmeasured "process noise" affecting the system state (and also models the inaccuracy of the cell model, to some extent). The system output is $y_k$ and $v_k$ is the measurement noise that usually models noise in sensors. The functions, $f(\ )$ and $g(\ )$ in the equations describe the (possibly nonlinear) behavior defined by the specific cell model employed.

To elaborate further, the system input vector $u_k$ usually comprises the instantaneous cell current $i_k$. It may also encompass additional information such as the cell temperature $T_k$, an estimate of the cell's capacity $C$, and/or an estimate of the cell's internal resistance $R_k$, for example. The system output is typically a scalar but can also be a vector. In this context, we consider the output as the loaded terminal voltage of the cell, not its at-rest open circuit voltage (OCV). The system's state vector $x_k$ in a summarized manner, represents the cumulative effect of all past inputs to the system, enabling the prediction of the present output solely based on the state and present input. Knowledge of past input values is not necessary. Our method mandates the inclusion of State of Charge (SOC) as one component within the state vector, as described in Section 3.

Numerous cell models have been proposed in various studies, serving different purposes. In Section 2, we provide an overview of a few of these models. Our specific focus is on modeling cell dynamics for state-of-charge (SOC) estimation in autonomous electric vehicle (AEV) battery packs. The AEV application poses significant challenges, including demanding rate requirements of approximately ±25C, highly dynamic rate profiles, and operating temperatures ranging from -30°C to 50°C. This stands in contrast to relatively fewer demanding scenarios like portable electronic applications with constant power output and fractional C rates. It is worth noting that methods for cell modeling and SOC estimation that perform well in portable electronic devices often fall short in the AEV application. Thus, achieving precise SOC estimation in an AEV necessitates a highly accurate cell model. The cells considered in this paper are Lithium-Ion Polymer-based, developed collaboratively by Compact Power Inc. in the United States and LG Chem, Ltd. in Daejeon, Korea. These cells have a nominal capacity of approximately 8Ah and are optimized for power-intensive applications. The approach presented in this paper offers a remarkably precise representation of the dynamics exhibited by these cells. Moreover, the method exhibits a high level of generality, making it applicable to a wide range of battery systems with varying chemistries and applications.

The structure of this paper is as follows: First, we provide a concise review of SOC estimation methods that incorporate companion cell modeling approaches. Next, we elucidate how our approach differs, emphasizing the essential requirement of including SOC as a component of the system state, and highlighting the advantages associated with this choice. Subsequently, we propose potential structures for the cell model and outline methods for determining the model parameters. We also detail the testing equipment, cells, and regimen employed for cell modeling. Finally, we evaluate the results obtained and draw conclusions based on our findings.

## II.  ALTERNATE METHODS FOR CELL MODELING AND SOC ESTIMATION

We begin by examining the existing literature to assess if the current methods align with our requirements. Our specific application involves modeling cell dynamics to estimate SOC in an AEV battery pack. We observe that many papers on cell modeling do not directly address SOC estimation, while many papers on SOC estimation include some discussion on cell modeling. Consequently, several references cited in this paper are focused on SOC estimation. For a comprehensive overview of these methods in greater detail than presented here, we refer readers to reference [2].

In our intended application, it is crucial for the cell model to exhibit accuracy across all operating conditions. These conditions encompass high rates, where many papers consider rates up to approximately ±1C for portable electronic applications, whereas we need to consider rates up to about ±25C. Furthermore, the model must account for temperature variations within the automotive range of -30°C to 50°C and accommodate highly dynamic rates, distinguishing it from the comparatively milder conditions encountered in portable electronics and battery electric vehicle applications. It is also essential to consider charging (regen) in the method.

Additionally, we require non-invasive techniques that rely solely on readily available signals. This requirement arises from the AEV environment, where the battery management system (BMS) lacks direct control over the current and voltage experienced by the battery pack, as this falls within the domain of the vehicle controller and inverter. Therefore, we must rely on measurements such as instantaneous cell terminal voltage, cell current, and cell external temperature. Furthermore, our choice of cell chemistry restricts the range of approaches we can consider. Techniques specific to lead-acid chemistries, for example, are not suitable for LiPB cells.

## A. *Laboratory and Chemistry-Dependent Methods*

Several methods are not applicable in our specific application. Firstly, a laboratory method that involves completely discharging a cell to determine its remaining capacity cannot be employed in the AEV context. This approach is both unfeasible and counter-productive for our purposes. Secondly, chemistry-dependent techniques designed for lead-acid batteries, such as the Coup de Fouet measurement or the measurement of electrolyte physical properties, are unsuitable since our application employs LiPB cells. Lastly, open-circuit voltage (OCV) measurements are not practical for dynamic SOC estimation. While OCV can be used to infer SOC by referencing a lookup table, this method requires extended periods of battery inactivity, sometimes spanning hours, for the terminal voltage to approach OCV. Hence, this approach is impractical for our dynamic SOC estimation requirements.

## B. *Electro-chemical Modeling*

One approach to modeling cell electrical dynamics involves a meticulous consideration of the chemical reactions and processes taking place within the cell. This approach delves into the specific reactions occurring at the anode and cathode, as well as the ion transfer process within the electrolyte. These models, as exemplified in reference [3], can yield highly accurate predictions of terminal voltage. However, extracting SOC directly from these models poses a challenge, and it would be arduous (if even possible) to measure the numerous physical parameters required on a per-cell basis, particularly in high-volume consumer products. Consequently, we have not pursued this particular approach.

## C. *Impedance Spectroscopy*

Another broad category of cell modeling involves the measurement of cell impedances across a wide range of AC frequencies [4-8]. Typically, an equivalent circuit model of the cell is constructed using resistors, capacitors, inductors, and/or complex impedances. The values of the model parameters are determined through least squares fitting based on measured impedance values. SOC is typically considered as an input to the model, as cell impedance exhibits a dependency on SOC. Consequently, SOC can be indirectly estimated by measuring cell impedance and establishing correlations with known impedances at different SOC levels.

However, we must disregard this method for our specific application, as we lack a direct means to introduce signals into cells for impedance measurements. We rely on the vehicle to generate and dissipate all the energy flowing through the battery pack, leaving us unable to inject specific signals. Although the impedances could potentially be generated using a fast Fourier transform (FFT) approach, utilizing available measurements as $Z(e^{j\omega}) = E(e^{j\omega})/I(e^{j\omega})$, we would need to ensure that the current signal $i(t)$ is persistently exciting and that $I(e^{j\omega})$ does not have any zero values. This guarantee would be violated, for instance, if the battery pack remained inactive for a certain period, which is a common occurrence. Moreover, depending on the block length of the FFT, this method could introduce an unacceptable time delay in measuring impedance and consequently estimating SOC.

## D. Circuit Models

A number of papers propose equivalent circuit models for cells [9-12]. These models commonly incorporate a high-valued capacitor to represent the open-circuit voltage (OCV), while the rest of the circuit represents the cell's internal resistance and dynamic effects like terminal voltage relaxation. SOC can be inferred from the OCV estimate through table lookup. Both linear and nonlinear circuit models can be employed for this purpose. However, our findings indicate that linear circuit models do not achieve the desired level of performance.

## E. Coulomb Counting

The final method discussed in the literature focuses on SOC estimation directly through Coulomb counting. This can be done in an "open-loop" fashion, which is often imprecise due to sensor errors, or a more accurate "closed-loop" approach. The feedback mechanism can be designed empirically [13] or employ a theoretically justified method such as Kalman Filtering [14-15] to generate the feedback. All the Kalman filtering-based methods described in the literature (that we are aware of) utilize a circuit model of the cell where capacitor voltages represent OCV and relaxation effects, enabling the estimation of OCV and subsequent inference of SOC.

Our approach also utilizes the Kalman filtering method. However, the fundamental aspect that distinguishes our model from those reported in the literature is that SOC is directly considered as a state of the system. This approach offers a significant advantage in that the Kalman filter provides a dynamic estimate of SOC and its uncertainty. This concept is discussed in more detail in [1]. Instead of providing a single SOC value to the vehicle controller at a particular time (e.g., "about 55%"), our algorithm is capable of reporting that the SOC is 55%±7%, for instance. This enables the vehicle controller to confidently utilize the battery pack's full operating range without concerns of over- or under-charging cells.

## III. MODEL STRUCTURES

In order to use the Kalman methods we propose to estimate SOC, the cell model must be represented in discrete-time state-space form. Specifically, we assume the form of equations (1). The difference between the models, then, depends on the definitions of $x_k, u_k, f(\ )$ and $g(\ )$. We also require that SOC is a member of the state vector. To be complete, we give a list of definitions culminating in a careful definition of SOC.

**Definition:** The cell high operational voltage limit is called $v_h$. Here, we may use $v_h = 4.2$ V.
**Definition:** The cell low operational voltage limit is called $v_l$. Here, we may use $v_l = 3.0$ V.
**Definition:** A cell is fully charged when its voltage reaches $v = v_h$ after being charged at infinitesimal current levels.
**Definition:** A cell is fully discharged when its voltage reaches $v = v_l$ after being drained at infinitesimal current levels.
**Definition:** The capacity of a cell is the maximum number of Ampere-hours that can be drawn from the cell before it is fully discharged, at room temperature (25°C), starting with the cell fully charged. Definition: The nominal capacity of the cell is the number of Ampere-hours that can be drawn from the cell at room temperature at the $C/40$ rate, starting with the cell fully charged.
**Definition:** The SOC of the cell is the ratio of the remaining capacity to the nominal capacity of the cell, where the remaining capacity is the number of amp-hours that can be drawn from the cell at room temperature at the $C/40$ rate.

With these definitions in place, we can then investigate some mathematical relations involving SOC. Particularly:

$$SOC(t) = SOC(0) - \int_0^t \frac{\eta(i(\tau))i(\tau)}{C} d\tau \qquad (2)$$

where $C$ is the nominal capacity of the cell, $i(t)$ is the cell current at time $t$, and $\eta(i(t))$ is the Coulombic efficiency of the cell. (Here, we use $\eta(i(t)) = 1$ for discharge and $\eta(i(t)) = 0.995$ for charge).

A discrete time approximate recurrence may then be written as:

$$SOC_{k+1} = SOC_k - \frac{\eta(i_k)i_k \Delta t}{C} \qquad (3)$$

where $\Delta t$ is the sampling period (in hours). Equation (3) is the basis for including SOC in the state vector of the cell model as it is in state equation format already, with SOC as the state and $i_k$ as the input. Our cell models will then be differentiated by the additional components in the state vector and the functional form of $f(\ )$ and $g(\ )$.

### A. Models with a Single State

We will first investigate models with a single state, i.e., SOC. These models share a common process equation (3). The difference between them is then the output equation. Several different forms are suggested in reference [16].

**Shepherd model:** $\qquad y_k = 4.2 - Ri_k - K_i/SOC_k \qquad (4)$
**Unnewehr universal model:** $\qquad y_k = 4.2 - Ri_k - K_i SOC_k \qquad (5)$
**Nernst model:** $\qquad y_k = 4.2 - Ri_k + K_1 \ln(SOC_k) \qquad (6)$
**Modified Nernst model:** $\qquad y_k = 4.2 - Ri_k + K_2 \ln(SOC_k) + K_3 \ln(1 - SOC_k) \quad (7)$

In these models, $y_k$ is the cell terminal voltage, $R$ is the cell internal resistance (different values may be used for charge/discharge and at different SOC levels if desired), $K_i$ is the polarization resistance and $K_1, K_2$, and $K_3$ are constants chosen to make the model fit the data well. The "modified Nernst" model of (7) reflects an additional term that we added to the Nernst model to cause it to fit our data better. All of the terms of (4) through (7) may be collected to make a "combined model" that performs better than any of the individual models alone.

**Combined model:**

$$y_k = K_0 - Ri_k - K_1/SOC_k - K_2 SOC_k + K_3 \ln(SOC_k) + K_4 \ln(1 - SOC_k) \qquad (8)$$

The unknown quantities in (8) may are estimated using a system identification procedure. This model has the advantage of being "linear in the parameters"; that is, the unknowns occur linearly in the output equation.

A simple way to find the parameters is then as follows: First form the vector $Y = [y_1, y_2, \ldots y_N]^T$ and the matrix $H = [h_1^T, h_2^T, \ldots h_N^T]^T$. The rows of $H$ are (transposes of) $h_j = [1, i_j^+, i_j^-, 1/SOC_j, SOC_j, \ln(SOC_j), \ln(1 - SOC_j)]^T$, where $i_j^+$ is equal to $i_j$ if $i_j > 0$, $i_j^-$ is equal to $i_j$ if $i_j < 0$, else $i_j^+$ and $i_j^-$ are zero.

Then,

$$Y = H\theta, \tag{9}$$

where $\theta = [K_0, R^+, R^-, K_1, K_2, K_3, K_4]^T$ is the vector of unknown parameters. The least-squares solution for $\theta$ is:

$$\theta = (H^T H)^{-1} H^T Y \tag{10}$$

This may be evaluated in MATLAB, for example, as theta $a = H \setminus Y$;

## B. Models with Multiple States to Track Relaxation

The combined model presented in equation (8) can be identified and implemented swiftly. However, a significant drawback of this model is its omission of any description of cell relaxation. Considering the necessity for accurate prediction of cell behavior in a dynamic AEV environment, it becomes essential to incorporate relaxation effects.

In a state-variable model, the dynamics are described by the state equation (1). To account for relaxation effects, we need to expand the state vector by introducing additional filter states. In our approach, we opt to implement filtered versions of SOC and the input current. This leads to the following augmented model:

$$x_{k+1} = \begin{bmatrix} 1 & 0 & 0 & 0 \\ w_1 & w_2 & 0 & 0 \\ 0 & 0 & w_4 & w_5 \\ 0 & 0 & -w_5 & w_4 \end{bmatrix} x_k + \begin{bmatrix} -1 & 0 \\ 0 & 1 \\ 0 & 0 \\ 1 & 0 \end{bmatrix} \begin{bmatrix} I_k^{\text{mod}} \\ w_3 \end{bmatrix}$$

$$y_k = w_6 + w_7 I_k^{\text{mod}} + \frac{w_8}{x_{k,1} + w_9} + [w_{10} \quad 10 \quad w_{11} \quad w_{12}] x_k, \tag{11}$$

where $I_k^{\text{mod}} = \eta(i_k)|i_k|^n \Delta t / C_p$, $n$ is the Peukert exponent and $C_p$ is the Peukert capacity. The first state of $x_k$ (that is, $x_{k,1}$) is SOC, as before. The output $y_k$ is terminal voltage, as before. The parameters of the model are found by system identification using measured cell data. We found that the model was able to predict cell behavior best when different sets of parameters were used for different levels of input current $i_k$.

To estimate the state of a dynamic system, such as the cell SOC, we will employ a Kalman filter. A comprehensive explanation of this approach can be found in [1]. For a detailed understanding of the system identification procedure, it is recommended to consult this reference, which provides in-depth information on Kalman filtering. Additional insights can be gained from reference [17]. Moreover, we can utilize a Kalman filter for system identification purposes. In this case, the weights or parameters of the cell model are regarded as the state of a presumed "true" dynamic system:

$$\begin{aligned} W_{k+1} &= W_k \\ d_k &= y_k + v_k. \end{aligned} \tag{12}$$

Here, $W_k$ is the "truth/optimum" weight vector at time $k$ and has as components the weights $w_1$ through $w_{12}$. The optimum weight vector is constant, explaining the dynamics in the top line. The "output" of the optimum weight dynamics is the desired response, which is equal to the cell output plus the estimation error. We can create an extended Kalman filter to iteratively estimate the state (weight vector) of the cell model:

$$\begin{aligned}
\hat{W}_{k+1} &= \hat{W}_k + L_k(d_k - y_k) \\
L_k &= P_k C_k [C_k^T P_k C_k + R_k]^{-1} \\
P_{k+1} &= P_k - L_k C_k^T P_k.
\end{aligned} \quad (13)$$

Here, $P_k$ is the approximate conditional error covariance matrix, initialized to a diagonal matrix with small values, $R_k \leq 1$, and $C_k^T = dy_k/dW$.

To compute $dy_k/dW$ we first note that $y_k = fn(x_k, I_k^{\text{mod}}, W)$, $x_k = fn(x_{k-1}, I_{k-1}^{\text{mod}}, W)$ and use the chain rule for total differentials:

$$\begin{aligned}
\frac{dx_k}{dW} &= \frac{\partial x_k}{\partial W} + \underbrace{\frac{\partial x_k}{\partial x_{k-1}}}_{A_{k-1}} \frac{dx_{k-1}}{dW} + \frac{\partial x_k}{\partial I_{k-1}^{\text{mod}}} \underbrace{\frac{dI_{k-1}^{\text{mod}}}{dW}}_{0} \\
\frac{dy_k}{dW} &= \frac{\partial y_k}{\partial W} + \frac{\partial y_k}{\partial x_k} \frac{dx_k}{dW} + \frac{\partial y_k}{\partial I_{k-1}^{\text{mod}}} \underbrace{\frac{dI_{k-1}^{\text{mod}}}{dW}}_{0}
\end{aligned} \quad (14)$$

In the second line,

$$\frac{\partial y_k}{\partial W} = \begin{bmatrix} 0 & 0 & 0 & 0 & 0 & 1 & I_k^{\text{mod}} & \dfrac{1}{x_{k,1}+w_9} & \dfrac{-w_8}{(x_{k,1}+w_9)^2} & x_{k,1} & x_{k,3} & x_{k,4} \end{bmatrix},$$

$$\frac{\partial y_k}{\partial x_k} = \begin{bmatrix} w_{10} & 10 & w_{11} & w_{12} \end{bmatrix} + \begin{bmatrix} \dfrac{-w_8}{(x_{k,1}+w_9)^2} & 0 & 0 & 0 \end{bmatrix} = \begin{bmatrix} w_{10} - \dfrac{w_8}{(x_{k,1}+w_9)^2} & 10 & w_{11} & w_{12} \end{bmatrix},$$

(15)

and $dx_k/dW$ is computed in the first line. In the first line:

$$\frac{\partial x_k}{\partial W} = \begin{bmatrix} 0 & 0 & 0 & 0 & 0 & 0 & 0 & 0 & 0 & 0 & 0 & 0 \\ x_{k-1,1} & x_{k-1,2} & 1 & 0 & 0 & 0 & 0 & 0 & 0 & 0 & 0 & 0 \\ 0 & 0 & 0 & x_{k-1,3} & x_{k-1,4} & 0 & 0 & 0 & 0 & 0 & 0 & 0 \\ 0 & 0 & 0 & x_{k-1,4} & -x_{k-1,3} & 0 & 0 & 0 & 0 & 0 & 0 & 0 \end{bmatrix},$$

$$\frac{\partial x_k}{\partial x_{k-1}} = A_{k-1} = \begin{bmatrix} 1 & 0 & 0 & 0 \\ w_1 & w_2 & 0 & 0 \\ 0 & 0 & w_4 & w_5 \\ 0 & 0 & -w_5 & w_4 \end{bmatrix},$$

(16)

and $dx_{k-1}/dW$ is a previously computed and stored version of $dx_k/dW$. All terms are accounted for, and the algorithm is complete.

## C. Radial Basis Function Model

Adding linear filter states into the model enhances its predictive capability for cell behavior. However, as LiPB cells inherently exhibit nonlinear characteristics, further improvement can be achieved by employing a fully nonlinear dynamic cell model. To accomplish this, we will utilize radial-basis-function (RBF) networks in conjunction with a black-box system identification procedure.

An RBF network provides a localized approximation of the function it represents. It calculates its output by taking a weighted sum of (hyper) Gaussian shapes. More precisely, it computes the function as follows:

$$y_k = \sum_{j=1}^{N} w_j \exp\left(-\frac{1}{\sigma_j^2}\|u_k - t_j\|^2\right) + w_{N+1}, \tag{17}$$

where $N$ is the number of bases, $w_j$ is the weight connecting the $j$ th basis function to the output, $\sigma_j$ is the "standard deviation" or width parameter of the $j$ th basis function, $x_k$ is the vector input to the network, and $t_j$ is the center of the $j$ th basis function. Here, $u_k$ includes the states of the system: e.g., $x_k = [y_{k-1}, \text{SOC}_k]^T$ as well as the cell current $i_k$. Figure 1 depicts a cartoon illustrating the approximation of a smooth function using RBFs.

The red line represents the function being approximated. By combining two blue Gaussian shapes with distinct centers, widths, and heights, a close approximation to the red line is achieved. This concept extends to higher dimensions as well. During the training of an RBF, the objective is to identify the appropriate set of centers, widths, and output scales that effectively approximate the target function.

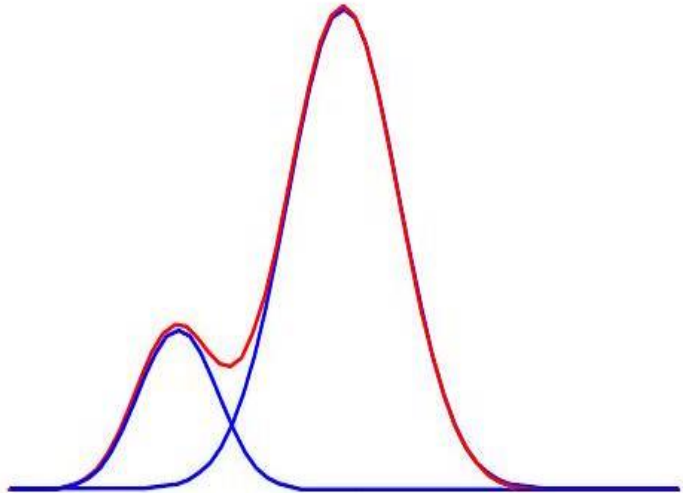

**Figure 1:** Cartoon illustrating how the function drawn as a red line may be approximated by the sum of two Gaussian shapes drawn as blue lines.

The parameters of an RBF network can be determined from data using a Kalman filter, following the same approach outlined in Section 3.2. However, the specific details of this process will not be elaborated upon here.

## IV. CELL TESTING AND MODEL FITTING RESULTS

To assess the performance of the proposed cell models in capturing the dynamics of LiPB cells, we conducted tests using prototype cells. The tests were carried out in a Tenny thermal chamber set at a temperature of 25∘C, with an Arbin cell cycler. Prior to the tests, the cells were fully charged. The test cycles consisted of pulsed discharge phases followed by rest intervals, and then pulsed charge phases with additional rest periods. Measurements, including voltage, current, and Ah discharged/charged, were recorded at one-second intervals. The collected data was utilized to identify the parameters of the three cell models. Subsequently, these models were employed to predict the terminal voltage for the conducted tests.

Figures 2-4 present a comparison between the model's predicted terminal voltage and the actual measured terminal voltage for three representative tests involving pulsed ±1C rates, pulsed ±2C rates, and pulsed ±4C rates. In each plot, the true cell voltage is depicted by the red line, while the model's prediction is illustrated by the blue line.

Figure 2 specifically focuses on the "combined model", highlighting a comparison between the measured data and the model's output. It should be noted that since this model lacks filter states, the prediction may lack relaxation effects.

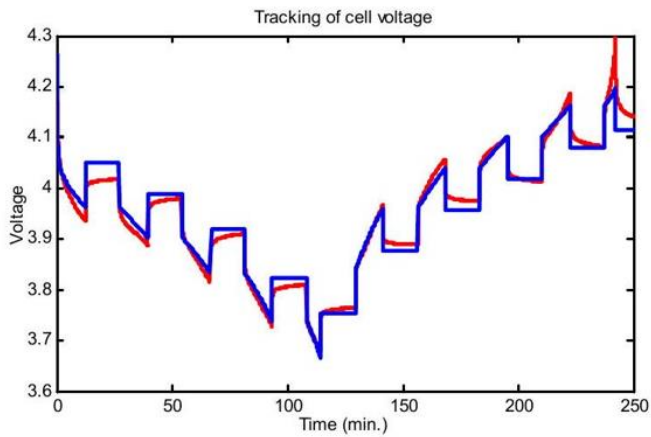

Pulsed current at $\pm 1C$ rates.

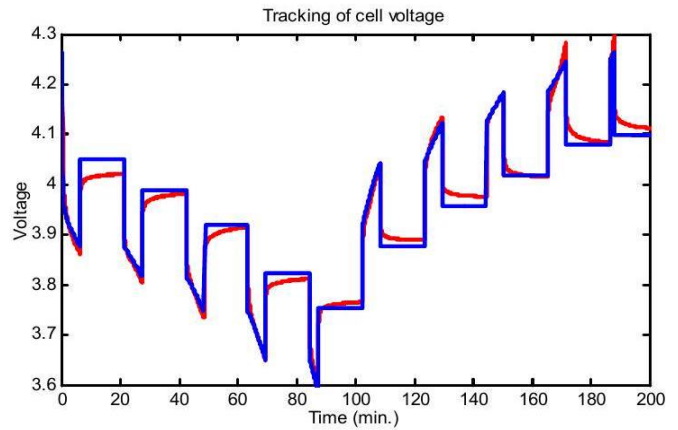

Pulsed current at ±2C rates.

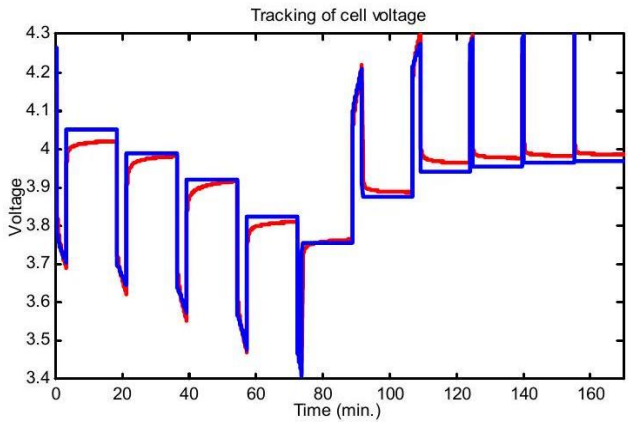

Pulsed current at ±4C rates.

**Figure 2:** Cell voltage tracking is achieved through the utilization of the single-state model. The red line represents the actual cell voltage, while the blue line represents the voltage predicted by the cell model. To conduct the tests, the cell was subjected to pulsed currents at rates of ±1C, ±2C, and ±4C, interspersed with rest periods.

Figure 3 shows results from the "filter state" model. It does a much better job of capturing the relaxation dynamics but is still noticeably flawed due to its nearly linear nature.

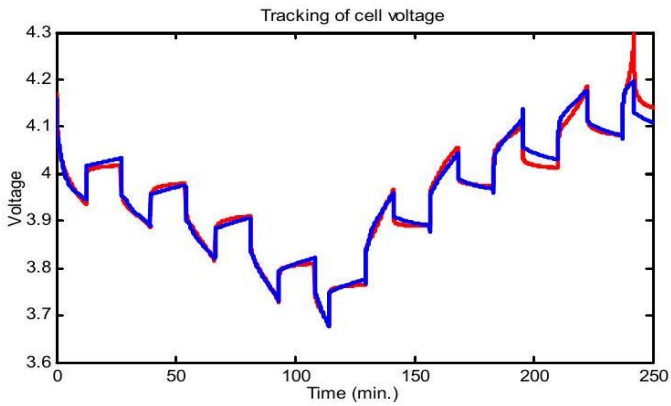

Pulsed current at ±1C rates.

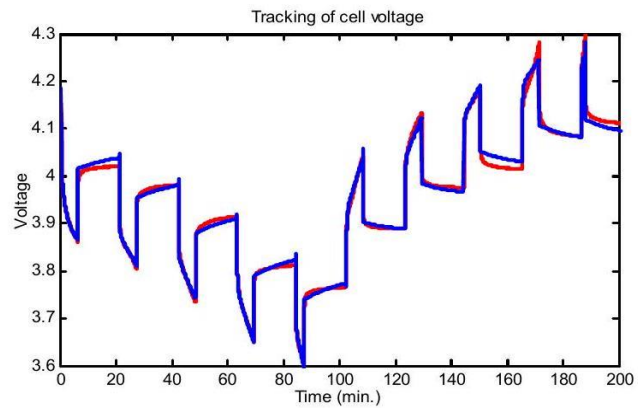

Pulsed current at ±2C rates.

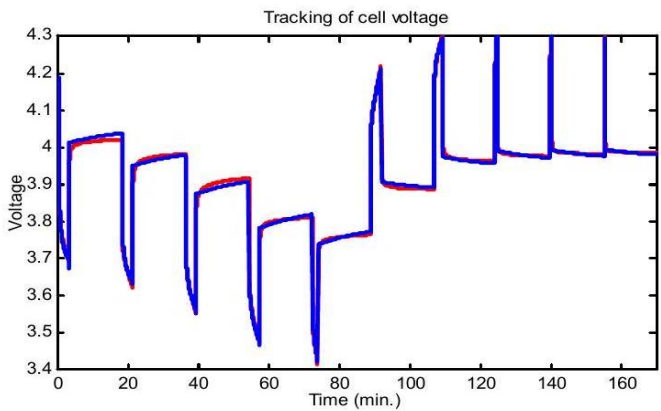

Pulsed current at ±4C rates.

**Figure 3:** To track cell voltage, the filter-state model is employed. The true cell voltage is depicted by the red line, while the voltage predicted by the cell model is represented by the blue line. The cell tests consisted of pulsed currents at rates of ±1C, ±2C, and ±4C, interspersed with rest periods.

Figure 4 displays the outcomes obtained from the "radial-basis-function model". It is evident that the model's output closely aligns with the actual cell output. This indicates that the model has effectively captured the cell's dynamics. In this particular case, a 100-RBF network was employed, resulting in an RMS estimation error of approximately 2mV. Notably, this error is lower than the expected quantization noise floor in our BMS implementation.

To further analyze the performance, Figure 5 showcases a plot illustrating the RMS estimation error of the model as a function of the number of RBF kernels utilized. The plot demonstrates that this approach allows for achieving arbitrary precision by increasing the number of RBFs until the desired level of accuracy is attained.

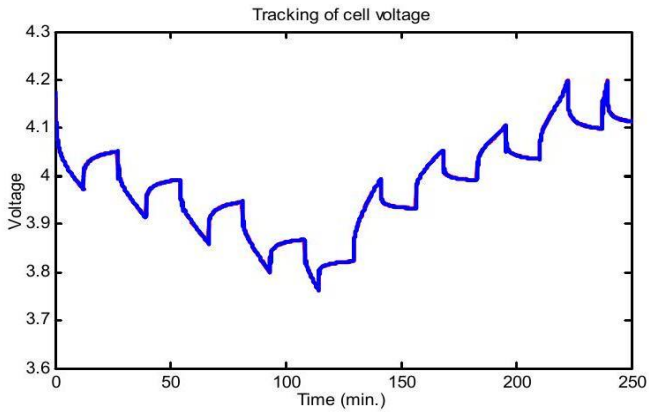
Pulsed current at ±1C rates.

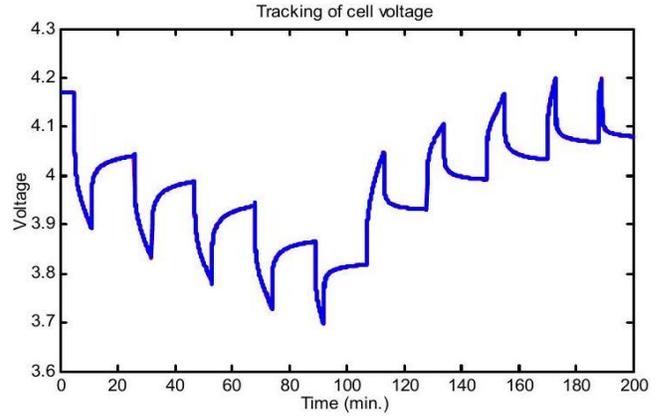
Pulsed current at ±2C rates.

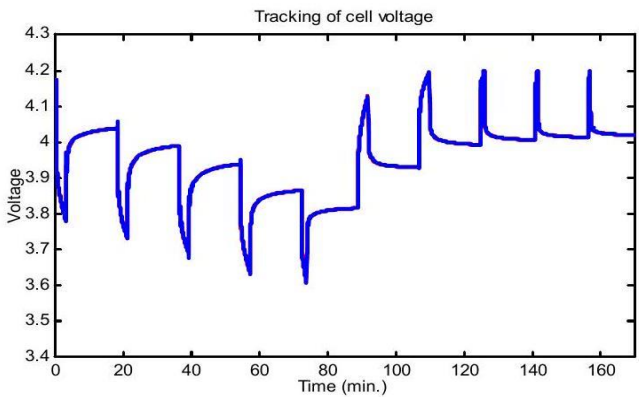
Pulsed current at ±4C rates.

**Figure 4:** Cell voltage tracking is accomplished using the radial-basis-function network model. The red line corresponds to the actual cell voltage, while the blue line represents the voltage predicted by the cell model. The cell tests involved pulsed currents at rates of ±1C, ±2C, and ±4C, with intermittent rest periods.

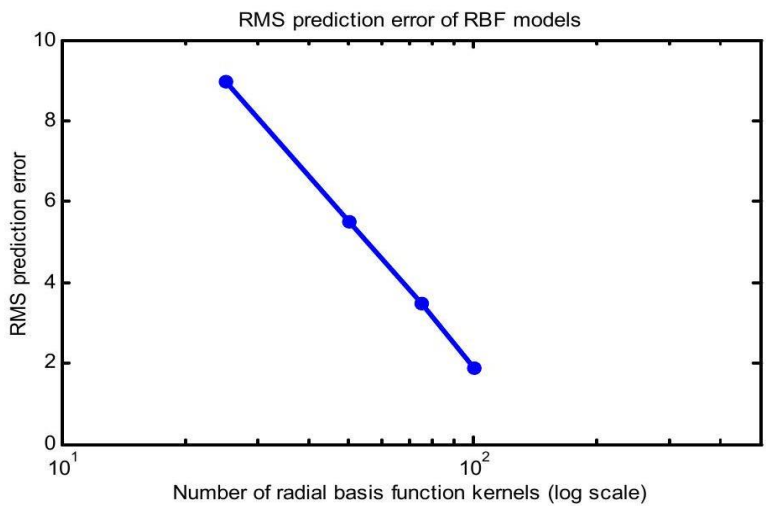

**Figure 5:** RMS prediction/modeling error using RBF networks with different numbers of basis functions (kernels).

Finally, Figure 6 shows results of a much more difficult modeling problem. Rather than simple pulsed charge/discharge cycles, it shows cell test results following a UDDS drive cycle, repeated a number of times over the SOC range of 0 to 1. An RBF network of the same size was used to identify this signal. Note that space does not permit lengthy discussion of model temperature dependence. Preliminary work indicates that temperature may be included as another input to the RBF input vector for accurate modeling over the required temperature range.

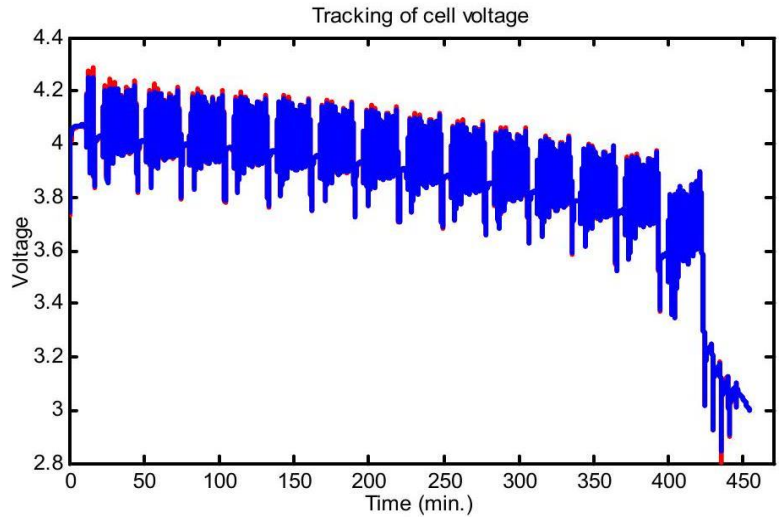

**Figure 6:** Tracking cell voltage in a very dynamic AEV test using the RBF model with 100 basis kernels.

## V.   CONCLUSIONS

In conclusion, this paper has introduced three mathematical state-space structures to model the dynamics of LiPB AEV cells, aiming to enable state-of-charge (SOC) estimation through Kalman filtering. Among the proposed models, the single-state model is the simplest but exhibits the lowest performance. On the other hand, incorporating filter states enhances performance at the expense of increased complexity. The final structure, utilizing radial-basis-function networks, offers a scalable complexity to effectively capture the dynamics, yielding the best performance among all models tested. Additionally, the SOC estimation results confirm the notion that "the better the model, the better the SOC estimation" [1]. Based on these findings, the RBF model emerges as the most favorable choice among the evaluated options.

## VI. BIOGRAPHY SECTION

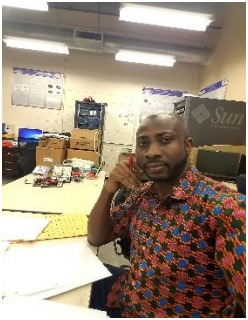

**Qasim Ajao** received his B.Sc. degree in Electrical Engineering from Yaba College of Technology in Lagos State in 2011. He later obtained his M.Sc. from Georgia Southern University in Georgia, United States, in 2019. Currently, he is pursuing a Ph.D. and working as a research scholar at Cardiff University, where he closely collaborates with Dr. A. R. Abdul Ameen at the National Institute of Technology in Cardiff, United Kingdom. His ongoing research focuses on several areas, including autonomous electric vehicles, Power Distributed and Energy Storage Systems, and Impact Research for STEM Development in Nigeria and Sub-Saharan African Nations.
E-mail: Qasim.ajao@ieee.org   URL: https://qasimajao.academia.edu/

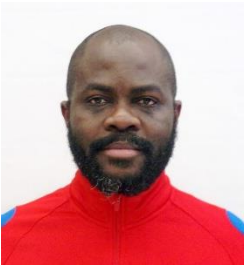

**Lanre G. Sadeeq** received the B.Sc. degree in Electrical Engineering from Yaba College of Technology in Lagos State in 2011, and the M.Sc. from Georgia Southern University in Georgia, United States, in 2020. He is currently a academic scholar and independent researcher in the United States. His current research interests include Power Protection Systems, Electric vehicles, Power Distributed and Energy Storage Systems, and with Impact focus on Research for STEM Development.
E-mail: 05lanre.s@gmail.com